\documentclass{emulateapj}

\begin{document}

\title{Absolute Calibration and Characterization of the Multiband Imaging
Photometer for Spitzer.  I.  The Stellar Calibrator Sample and the 24~\micron\
Calibration}

\shorttitle{MIPS 24~\micron\ Calibration}

\author{C.~W.\ Engelbracht\altaffilmark{1},
   M.\ Blaylock\altaffilmark{1},
   K.~Y.~L.\ Su\altaffilmark{1},
   J.\ Rho\altaffilmark{2},
   G.~H.\ Rieke\altaffilmark{1},
   J.\ Muzerolle\altaffilmark{1},
   D.~L.\ Padgett\altaffilmark{2},
   D.~C.\ Hines\altaffilmark{3},
   K.~D.\ Gordon\altaffilmark{1},
   D.\ Fadda\altaffilmark{2},
   A.\ Noriega-Crespo\altaffilmark{2},
   D.~M.\ Kelly\altaffilmark{1},
   W.~B.\ Latter\altaffilmark{4},
   J.~L.\ Hinz\altaffilmark{1},
   K.~A.\ Misselt\altaffilmark{1},
   J.~E.\ Morrison\altaffilmark{1},
   J.~A.\ Stansberry\altaffilmark{1},
   D.~L.\ Shupe\altaffilmark{2},
   S.\ Stolovy\altaffilmark{2},
   Wm.~A.\ Wheaton\altaffilmark{2},
   E.~T.\ Young\altaffilmark{1},
   G.\ Neugebauer\altaffilmark{1},
   S.\ Wachter\altaffilmark{2},
   P.~G.\ P\'{e}rez-Gonz\'{a}lez\altaffilmark{1,5},
   D.~T.\ Frayer\altaffilmark{2},
   and
   F.~R.\ Marleau\altaffilmark{2}
   }

\altaffiltext{1}{Steward Observatory, University of Arizona, Tucson, AZ 85721;
cengelbracht, mblaylock, ksu, grieke, jmuzerolle, kgordon, dkelly, jhinz,
kmisselt, jmorrison, jstansberry, eyoung, gxn@as.arizona.edu}
\altaffiltext{2}{Spitzer Science Center, 220-6, Caltech, Pasadena, CA 91125;
rho, dlp, fadda, alberto, shupe, stolovy, waw, wachter, frayer,
marleau@ipac.caltech.edu}
\altaffiltext{3}{Space Science Institute, 4750 Walnut Street, Suite 205,
Boulder, CO 80301; dhines@as.arizona.edu}
\altaffiltext{4}{NASA Herschel Science Center, Mail Code 100-22, California
Institute of Technology, 770 South Wilson Avenue, Pasadena, CA 91125;
latter@ipac.caltech.edu}
\altaffiltext{5}{current address:  Departamento de Astrof\'{\i}sica, Facultad
de CC.  F\'{\i}sicas, Universidad Complutense de Madrid, E-28040 Madrid,
Spain; pgperez@as.arizona.edu}

\begin{abstract}

We present the stellar calibrator sample and the conversion from instrumental
to physical units for the 24~\micron\ channel of the Multiband Imaging
Photometer for {\it Spitzer} (MIPS).  The primary calibrators are A stars, and
the calibration factor based on those stars is
$4.54\times10^{-2}$~MJy~sr$^{-1}$~(DN/s)$^{-1}$, with a nominal uncertainty of
2\%.  We discuss the data-reduction procedures required to attain this
accuracy;  without these procdures, the calibration factor obtained using the
automated pipeline at the {\it Spitzer} Science Center is 1.6\%$\pm$0.6\%
lower.  We extend this work to predict 24~\micron\ flux densities for a sample
of 238 stars which covers a larger range of flux densities and spectral types.
We present a total of 348 measurements of 141 stars at 24~\micron.  This
sample covers a factor of $\sim460$ in 24~\micron\ flux density, from 8.6~mJy
up to 4.0~Jy.  We show that the calibration is linear over that range with
respect to target flux and background level.  The calibration is based on
observations made using 3-second exposures; a preliminary analysis shows
that the calibration factor may be 1\% and 2\% lower for 10- and 30-second
exposures, respectively.  We also demonstrate that the calibration is very
stable: over the course of the mission, repeated measurements of our routine
calibrator, HD~159330, show a root-mean-square scatter of only 0.4\%.
Finally, we show that the point spread function (PSF) is well measured and
allows us to calibrate extended sources accurately; Infrared Astronomy
Satellite (IRAS) and MIPS measurements of a sample of nearby galaxies are
identical within the uncertainties.

\end{abstract}


\section{Introduction}
\label{sec:introduction}

Space-based infrared astronomy satellites have used a variety of methods to
perform flux calibration.  For example, the Infrared Astronomy Satellite
\citep[IRAS;][]{beichman88} extrapolated a ground-based 10~\micron\
calibration out to 60~\micron\ using stellar models and then used asteroids to
transfer the 60~\micron\ calibration to 100~\micron.  The Midcourse Space
Experiment (MSX) was calibrated in a set of bands from $\sim8-20~\micron$
using blackbodies ejected from the spacecraft itself \citep{price04}.  The
instruments aboard the Infrared Space Observatory (ISO) used a mix of
astronomical sources, from stars for the ISO Camera
\citep[ISOCAM;][]{blommaert03} to stars, asteroids, and planets for the ISO
Photometer \citep[ISOPHOT;][]{schulz02} and Short Wavelength Spectrometer
\citep[SWS;][]{decin00} and planets for the Long Wavelength Spectrometer
\citep[LWS;][]{gry03}.  The Diffuse Infrared Background Experiment (DIRBE)
imaged the sky in a series of bands from 1.25~\micron\ to 240~\micron\ and was
calibrated against stars, planets, and planetary nebulae \citep{hauser98}.
Like most of these missions, the instruments on board the {\it Spitzer} Space
Telescope \citep{werner04} are calibrated against celestial sources:  for
example, the Infrared Array Camera \citep[IRAC;][]{fazio04}, with photometric
bands from $\sim3-8~\micron$, is calibrated against stars \citep{reach05}, as
is the Infrared Spectrograph \citep[IRS;][]{houck04}.

The Multiband Imaging Photometer for Spitzer \citep[MIPS;][]{rieke04} has
three photometric bands at 24, 70, and 160~\micron.  Like the other {\it
Spitzer} instruments, the primary flux calibrators at 24 and 70~\micron\ are
stars; the calibrations are tied to the new infrared calibration by
\citet{rieke07} and are presented here (24~\micron) and in a companion paper
by \citet{gordon07} (70~\micron).  The sensitivity of the 160~\micron\ band is
sufficient to allow it to be calibrated against stars, too, but a strong,
short-wavelength ghost image limits the accuracy with which such hot sources
can be measured; hence, an asteroid-based transfer (and the color corrections
required to make the transfer) of the stellar calibration from the shorter
bands (similar in spirit to the IRAS 100~\micron\ calibration) is presented in
a companion paper by \citet{stansberry07}.

The MIPS 24~\micron\ calibration and capabilities can be compared to several
previous missions which overlap in wavelength.  For example, the quoted
absolute calibration uncertainties of the large surveys performed by IRAS at
25~\micron, by MSX at 21~\micron\ and by DIRBE at 25~\micron\ are 8\%
\citep{beichman88}, 7\% \citep{cohen01}, and 15\% \citep{hauser98},
respectively.  These missions were optimized for large area surveys and used
older detector technology. As a result, the beam size and point source
detection limits are poor compared to the {\it Spitzer} and ISO pointed
observatories.  Like MIPS, the ISOPHOT instrument was designed as a
general-user instrument, but the 25~\micron\ channel \citep[which achieved an
accuracy of $\sim10$\%;][]{klaas03} was an aperture photometer and not capable
of imaging.  Thus, MIPS is the first space instrument with an array detector
optimized for imaging near 24~\micron, achieving a point-source sensitivity of
$\sim60~\mu$Jy ($5\sigma$, 2000 seconds; {\it Spitzer} Observer's Manual),
with a beam of 6\arcsec\ FWHM.  As we show in this paper, the absolute
calibration of the MIPS 24~\micron\ channel is accurate to 2\%, which has
expanded the science possible in this wavelength range.  For example, the MIPS
24~\micron\ channel allows measurement of infrared excesses around stars with
lower effective temperatures and/or lower-mass debris disks than previously
possible \citep[e.g.,][]{rieke05,bryden06}.

Careful data reduction procedures are essential to reproduce the results
presented here; accordingly, we discuss the data reduction and photometry in
some depth in \S~\ref{sec:data}.  The calibration factor itself is computed in
\S~\ref{sec:calfactor}.  To explore the quality of the calibration, we expand
the sample of targets beyond those used to measure the calibration factor; the
details of the flux prediction are presented in \S~\ref{sec:sample}.  We use
the expanded sample in a series of tests of the calibration, discussed in
\S~\ref{sec:checks}.  Finally, we summarize our results in
\S\ref{sec:summary}.

\section{Data Reduction and Photometry}
\label{sec:data}

\subsection{Standard Processing}

\begin{figure}
\plotone{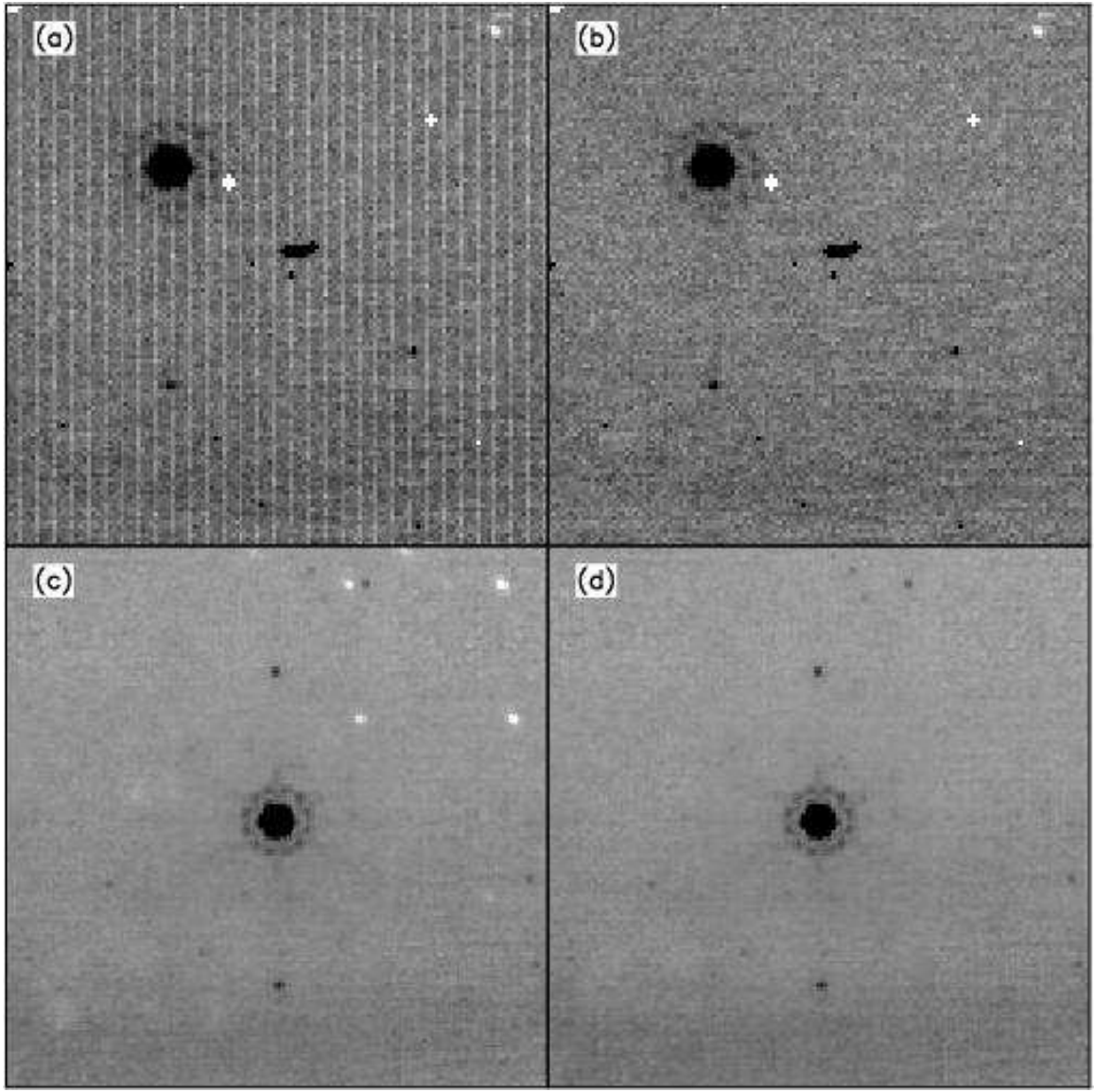}
\caption{Artifacts not discussed by \citet{gordon05} that are now fixed as
part of standard processing with the MIPS Data Analysis Tool (DAT),
illustrated using observations of HD~159330 (AORKEY 13587712) plotted in
reverse grayscale.  (a) The ``jailbar'' effect is most easily seen in
individual frames, here caused by a cosmic ray below and to the right of the
star, and (b) fixed as described in \S~{\ref{sec:data}}.  (c) The spots caused
by debris on the pick-off mirror (several sharp white spots above and to
the right of the star, as well as diffuse white regions below and to the left
of the star) are most easily seen in a mosaicked image and are (d) fixed
using separate flat fields for each scan mirror
position.\label{fig:std-artifacts}}
\end{figure}

The data were all obtained using the MIPS small-field photometry mode
astronomical observation template (AOT).  For most targets, 2 cycles of
photometry using 3-second exposures were obtained, resulting in 14
individual images at each of 2 telescope nod positions (excluding the
short exposure that starts the data-taking sequence at each of the two
telescope nod positions).  Starting with the raw data downloaded from the
{\it Spitzer} Science Center (SSC), these data were
processed using version 3.06 of the MIPS Data Analysis Tool
\citep[DAT;][]{gordon05}, which performs standard processing of infrared
detector array data (slope fitting, dark subtraction, linearity correction,
flat fielding, and mosaicking) as well as steps specific to the array used in
MIPS (droop correction and dynamic range extension using the first difference
frame)---these steps are described in more detail in that paper.  Experience
gained with the flight instrument has prompted the application of additional
processing, not discussed by \citet{gordon05}, to remove low-level
instrumental artifacts.  In particular, one of the four readouts on the array
can incur an offset of several counts per second (due to a bright source or
cosmic ray on detectors connected to that readout) relative to the other
three, resulting in a striping effect (dubbed ``jailbar'')---an additive
correction is applied to that readout to remove the effect.  Also, the
in-orbit flat field has several dark spots (due to particulate matter on the
pick-off mirror) that shift position on the array as the scan mirror (which
imposes small pointing adjustments as part of the normal observing sequence)
moves.  To remove these spots, a flat field specific to scan mirror position
is applied as part of the standard processing.  These artifacts are
illustrated in Figure~\ref{fig:std-artifacts}.

\subsection{Additional Processing}
\label{sec:addproc}

There are additional low-level artifacts present in the MIPS data after the
standard processing described above has been applied.  We assess the severity
of these effects and correct those that impact significantly our
measurements.  In roughly decreasing order of importance, these artifacts are:

The MIPS 24~\micron\ array is subject to medium-term ($\sim1$~week) gain
changes that affect pixels which have been exposed to highly saturating
sources, at the level of a few percent.  A typical MIPS 24~\micron\
observation will have several regions that have been affected this way as the
cumulative result of normal operations during that instrument campaign.  An
additional flat field, made from a median stack of the data with the source
masked out, is applied to remove the gain changes.

Background levels observed by MIPS at 24~\micron\ can change by several counts
per second from image to image throughout an observing sequence, likely due to
scattered zodiacal light that depends on the position of the MIPS scan mirror.
Uncorrected, this can affect the photometry of our faintest ($\sim10$~mJy)
calibrators at the level of about a percent.  To mitigate this effect, we
subtract the median level (after masking the source) from each image before
mosaicking.

Any source that falls on the MIPS 24~\micron\ array leaves a residual image of
about 0.5\% that decays with a timescale of roughly 10 seconds.  The dither
pattern used in the photometry AOT ensures that the source falls on a
different part of the array after every exposure, and thus any residual image
has decayed to a small fraction of a percent by the time the new source
position is within several full-width-half-maxima of a previous position.
Residual images have a negligible effect on the calibration factor we
derive here, and we perform no correction for them.

Sources that saturate the MIPS 24~\micron\ array in the full (3-second)
exposures used here can affect the offset of the pixels read out after the
ones on the source, and can even change the magnitude of the ``jailbar''
effect on those pixels, resulting in a different striping pattern above and
below the source on the array.  The offset is a few counts per second and only
occurs for the brightest sources, so it has a negligible effect on the
photometry of those sources and we make no correction here.

\begin{figure}
\plotone{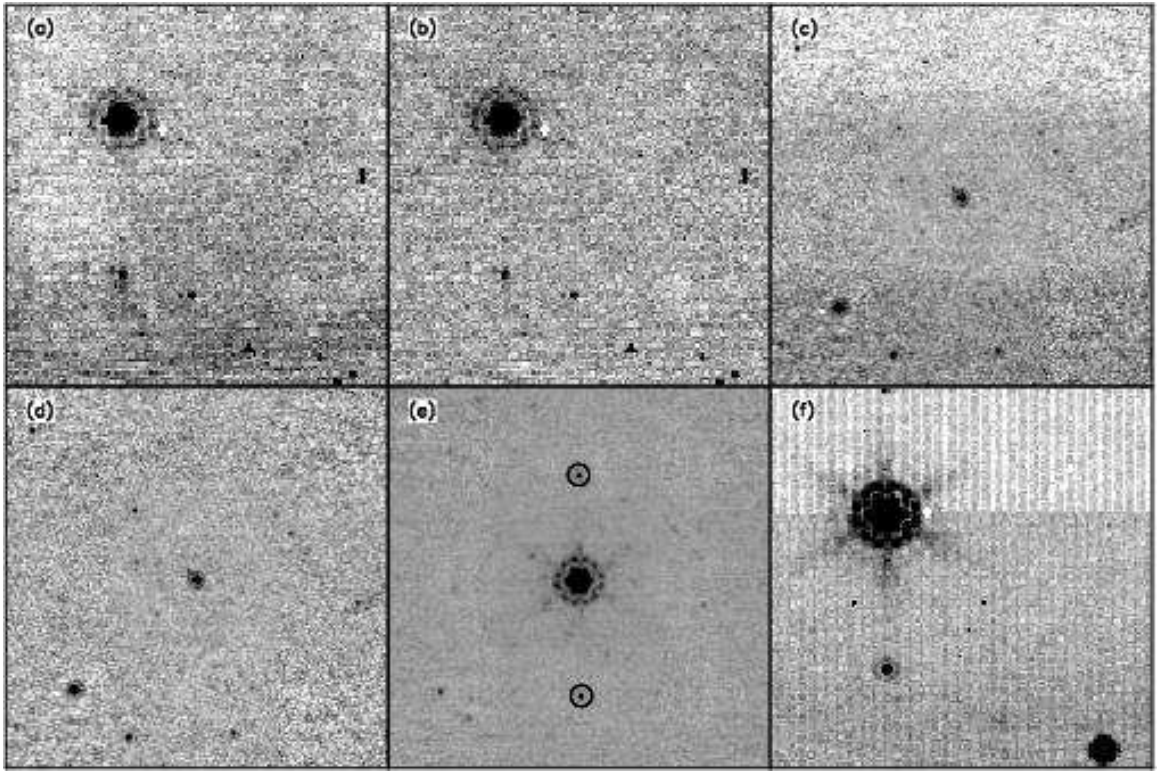}
\caption{Artifacts not removed by standard processing with the MIPS Data
Analysis Tool (DAT), plotted in reverse grayscale.  As discussed in
\S~\ref{sec:data}, (a) gain changes imposed by previous observations
(illustrated using a single-frame observation of HD~159330, AORKEY 12195328)
are (b) removed using a second flat field.  (c) Background changes as a
function of scan mirror position (illustrated using a mosaic observation of
HD~106965, AORKEY 13201920) are (d) removed before mosaicking.  For the
targets discussed here, the effects of (e) residual images (circled, illustrated
using a mosaic observation of HD~159330, AORKEY 12195328) and (f) sources
which saturate in 3 seconds (illustrated using a single-frame observation of
HD~180711, AORKEY 9805568) are small and we make no
correction.\label{fig:add-artifacts}}
\end{figure}

Examples showing the artifacts discussed here are shown in
Figure~\ref{fig:add-artifacts}.

\subsection{Mosaicking}

After applying the corrections discussed above, we coadd the data to make
a mosaic image of each source.  As the natural unit of a flat-fielded image is
surface brightness, the mosaicking step conserves surface brightness as it
removes optical distortions. The final images have square pixels $2\farcs45$
on a side, very similar to the natural pixel width near the center of the
array, $2\farcs49$.  The source positions in the photometry AOT range over the
central 2.5\arcmin\ of the array; to ensure that there are no calibration
changes across the full array, we have tested the distortion correction on
individual images.  Using the ``phot'' task in the Image Reduction and
Analysis Facility (IRAF\footnote{IRAF is distributed by the National Optical
Astronomy Observatories, which are operated by the Association of Universities
for Research in Astronomy, Inc., under cooperative agreement with the National
Science Foundation.}), we compared photometry in a fixed aperture (6 pixels in
radius, with a background annulus from $7-12$ pixels) for 416 measurements of
HD~042082 made during a focal plane survey which positioned the star over the
full extent of the array.  Without correction for distortion, the measured
counts increased smoothly by 12\% from the upper-left-hand corner of the array
to the lower-right-hand corner of the array, inversely proportional to the
12\% change in pixel area, from 6.91 square arcseconds to 6.11 square
arcseconds.  As expected, correcting the images for distortion eliminates this
trend, resulting in a $1\sigma$ dispersion in the measurements of only 0.8\%.

\subsection{Photometry}
\label{sec:photometry}

\begin{figure}
\plottwo{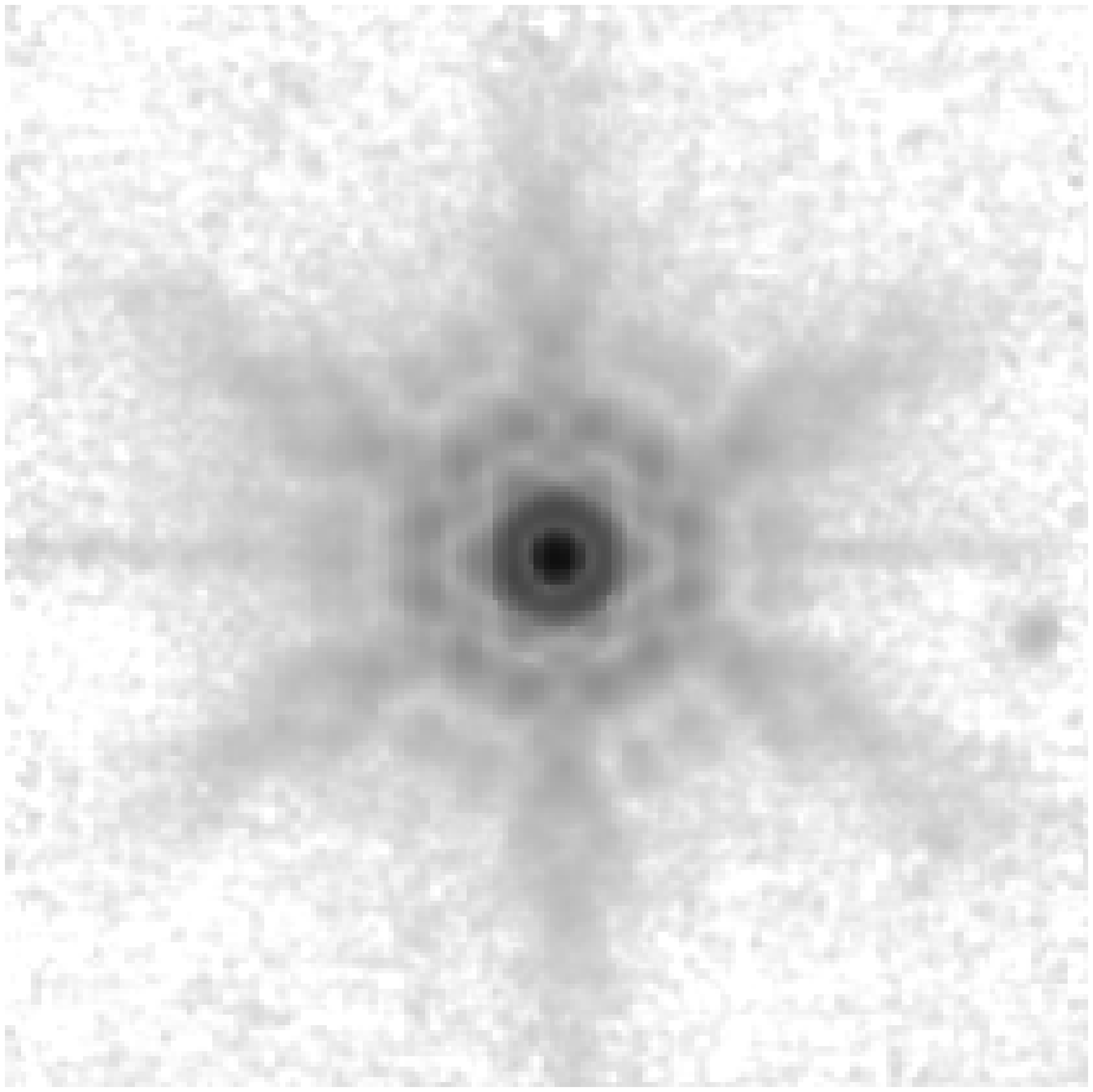}{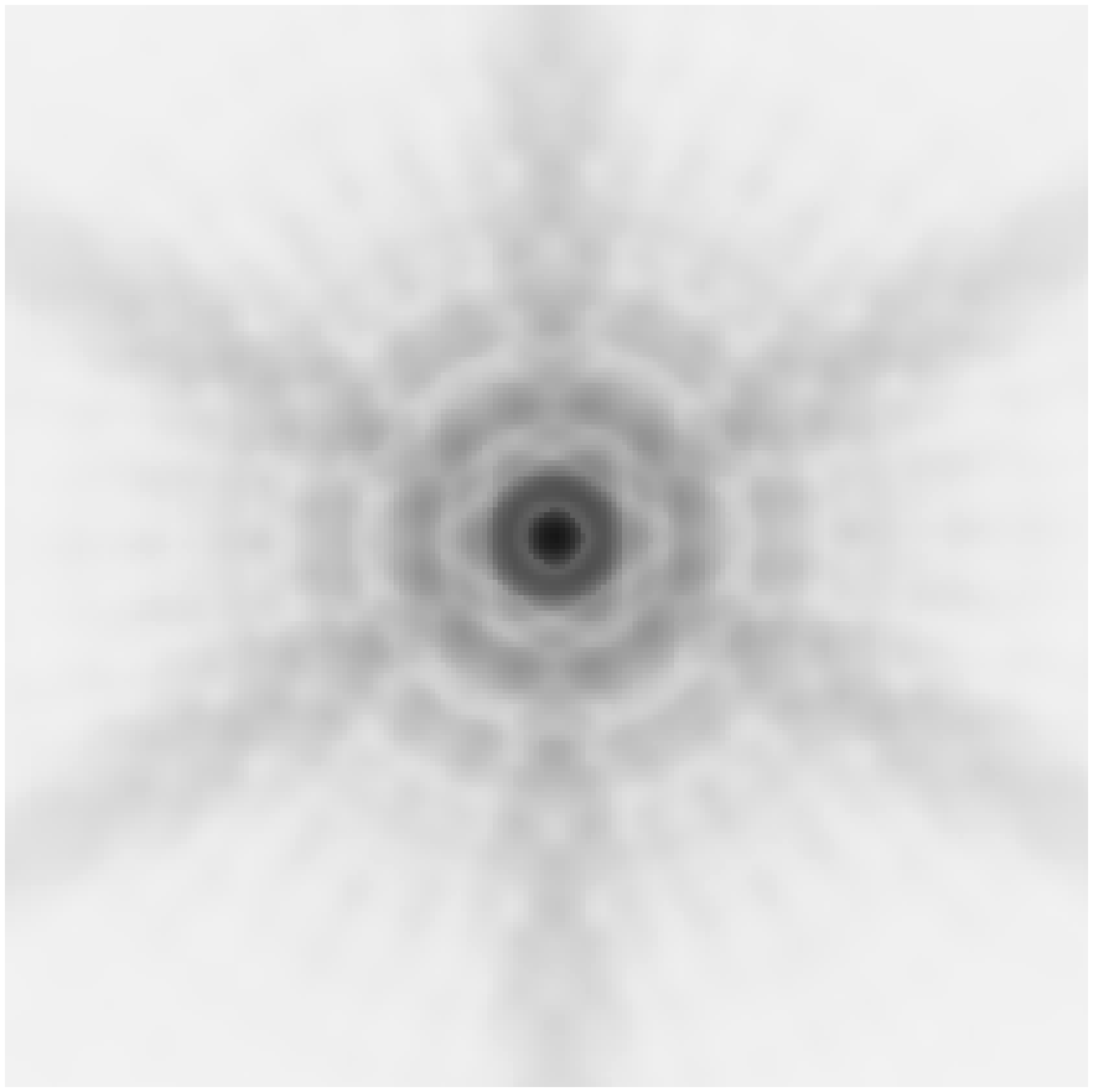}
\caption{Comparison of observed and model PSF, plotted in reverse grayscale.
On the left is shown an image of HD~009927; the horizontal feature through the
center of the image is a detector artifact.  The dynamic range (from the
peak brightness of 60,000 DN/s to the $1\sigma$ noise level of 3 DN/s) is
20,000.  The image on the right is a model generated by StinyTim, after
processing through the MIPS simulator (see \S~\ref{sec:photometry}).  Both
images have been heavily compressed using an asinh transform to show faint
structure and the grayscale levels have been adjusted by eye to match each
other.  \label{fig:psfcomp}}
\end{figure}

\begin{figure}
\plotone{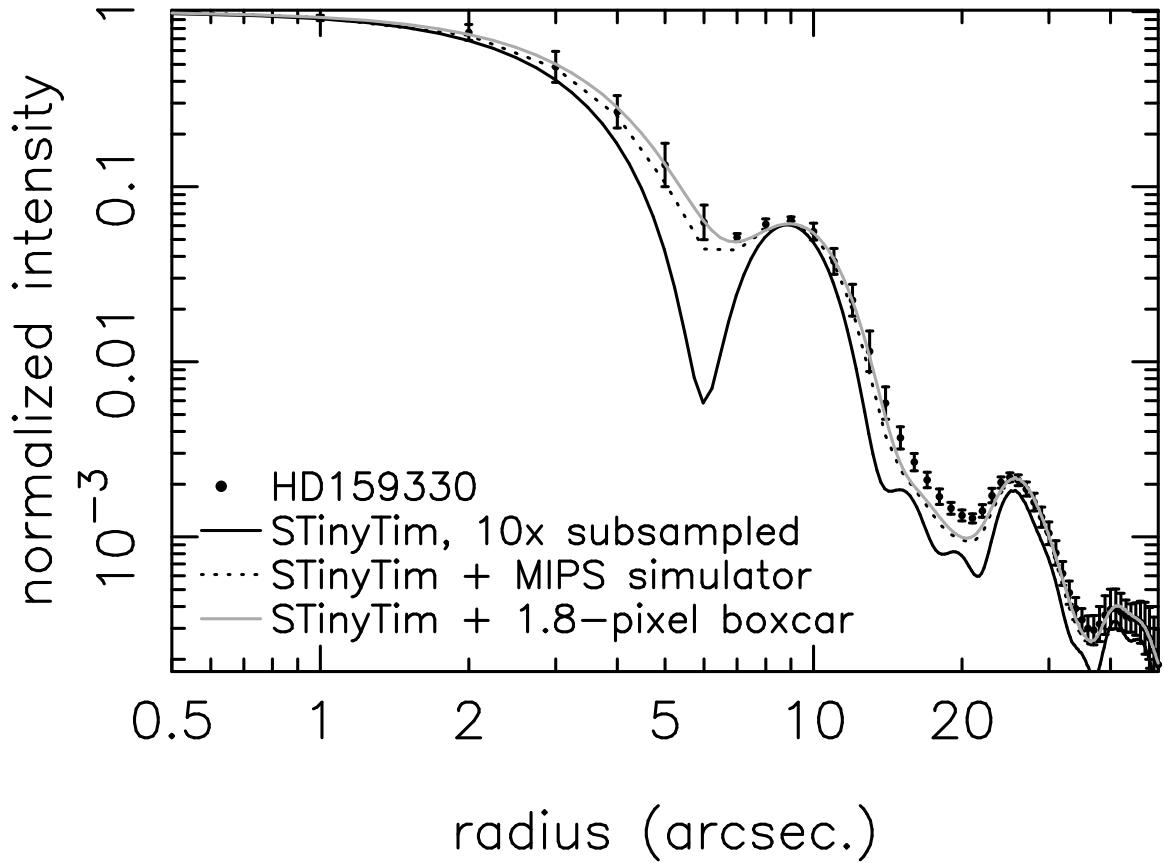}
\caption{The radial profile of a star (points), compared to a $10\times$
subsampled model profile generated by STinyTim (dark solid line), along with
that same model profile run through the MIPS simulator (dark dotted line; see
\S\ref{sec:data} for details) or smoothed by a 1.8-pixel boxcar (light solid
line).\label{fig:profile}}
\end{figure}

To compute aperture corrections for photometry, we make use of a point
response function (PRF) constructed as follows.  We start with a point spread
function (PSF) of a 10000~K blackbody (the spectral energy distributions of
the stars we use to calibrate differ negligibly from this at 24~\micron)
generated using the Spitzer TinyTim software \citep[STinyTim;][]{krist02}.
The PSF is computed at the center of the array using an image size
(10\arcmin\ width) large enough to encompass $>99$\% of the light.  We
``observe'' the model using the MIPS simulator (software designed to simulate
MIPS data, including optical distortions, using the same observing templates
used in flight) and mosaic the simulated data using the same software with
which we processed the real data.  We compare the result to one of the
calibration stars in Figure~\ref{fig:psfcomp}.  The structure predicted by the
model, down to very faint levels and in the diffraction spikes, is very
similar to what is observed on a real star.  Additional insight is gained from
a comparison of radial profiles (Figure~\ref{fig:profile}), where it can be
seen that the model is an excellent description of the data, out to the third
Airy ring.  A similar, and slightly improved, fit to the radial profile can be
obtained by smoothing the STinyTim model by the equivalent of a boxcar 1.8
pixels in width (possibly indicating a small amount of pointing jitter or
scattered light, neither of which is modeled by the simulator).  We apply the
``phot'' task in IRAF to the boxcar-smoothed image to compute the aperture
corrections, representative values of which are shown in
Table~\ref{tab:aperture-corrections}.  We find that the counts measured by
various photometry routines vary by about a percent, so we assign a 1\%
uncertainty to the aperture correction.

We perform aperture photometry on the observations using an aperture
35\arcsec\ in radius, with a background annulus from 40\arcsec\ to 50\arcsec\
in radius.  This aperture is large enough to minimize uncertainties due to
centroiding errors, and to ensure that any uncertainties in the aperture
correction have a small effect on the derived counts.  We derive the
uncertainty on each measurement from the scatter of the pixel values in the
background aperture.  The measurements are listed in
Table~\ref{tab:measurements}.

\subsection{Comparison to SSC Pipeline}
\label{sec:comparison}

We compare the measurements made on data reduced using the DAT to those
obtained using identical photometry procedures on data processed using the
automated pipeline at the SSC, which does not apply the second flat field
(made using a median stack of the data; see \S~\ref{sec:addproc})
or the background subtraction.  We convert the units of the SSC data
products from MJy/sr to DN/s by dividing by the FLUXCONV value found in the
header. Based on 90 measurements of HD~159330 processed using the 
S15.3.0 version of the pipeline, we find that the count rate in the
pipeline-reduced data is 1.6\% $\pm$ 0.6\% higher than in data
reduced using the DAT.  The difference between the reductions is
marginally significant ($2.7\sigma$) and may be due to the extra processing
steps applied to the DAT-processed data.

\subsection{Scan Map Calibration}

The results we present here rely on measurements of calibrator stars performed
using the photometry AOT, which images the sky in a nod-and-stare pattern.
MIPS has a second imaging mode, the scan map AOT, which images the sky using a
continuous track of the spacecraft.  The scan mode uses the same optical train
as the photometry mode and so we expect that the calibration derived here
would apply equally to both modes.  To check whether subtle differences exist
in the shape of the PSF that might affect the calibration, we compare
photometry in scan and photometry modes on two stars observed in multiple
epochs, HD~159330 and HD~163588.  We find that the radial profiles in scan and
photometry modes are identical within the uncertainties and that photometry in
these two modes agrees within $\sim1$\%, confirming that the photometry
calibration is applicable to scan-mode data.  Similar results were found by
\citet{fadda06}.

\section{The 24~\micron\ Calibration Factor}
\label{sec:calfactor}

Following \citet{rieke07}, we base the flux calibration of the MIPS
24~\micron\ band on A stars.  We adopt the sample of 22 A stars from
\citet{rieke07}, as a consistent suite of measurements is available for these
targets and the sample has already been vetted for sources which would bias
the calibration, such as those with an infrared excess.  Here and throughout
this work, monochromatic flux densities are computed at 23.675~\micron, the
effective wavelength of the 24~\micron\ band, although we will continue to use
the shorthand ``24~\micron'' for convenience.  We compute 24~\micron\ flux
densities using the extinction-corrected $K_s$ magnitudes computed by
\citet{rieke07}, a $K_s - [24]$ color difference of 0 magnitudes, and the
24~\micron\ (i.e., 23.675~\micron) zero point derived by
\citet{rieke07}, 7.17~Jy.  We apply the aperture correction derived in
\S~\ref{sec:data} (1.08) to the measurements from
Table~\ref{tab:measurements}, averaging multiple measurements when available.
The counts in that table are integrations over an aperture, so we
convert the implied unit of ``pixel'' (that we have ignored thus far) to an
angular area using the pixel size ($2\farcs45$ on a side) of our mosaics.  The
calibration factor is the weighted average of the ratio of the 24~\micron\
predictions to the observed count rate, or
$4.54\times10^{-2}$~MJy~sr$^{-1}$~(DN/s)$^{-1}$, with a formal uncertainty of
0.7\%.  At the pixel scale of our mosaics, this factor is equivalent to
$6.40\times10^{-6}$~Jy~pixel$^{-1}$~(DN/s)$^{-1}$ (which can be converted to
other pixels scales by multiplying by the square of the ratio of the pixel
size to 2\farcs45).  Following \citet{rieke07}, we assign an uncertainty of
2\% to the calibration factor to allow for systematic effects in propagating
the near-infrared measurements to 24~\micron.  Where care is taken to apply
the corrections discussed in \S~\ref{sec:data} and to treat calibration stars
and other targets in a consistent manner, this calibration accuracy can be
maintained in science data.  The data used to compute the calibration factor
are summarized in Table~\ref{tab:cal-factor}.

The calibration factor was derived using stars, and no color corrections
were applied to the measurements used here.  Stars are relatively blue ($f_\nu
\propto \nu^{-2}$) at MIPS wavelengths, so color corrections are required to
calibrate sources with different spectral distributions.  As shown by
\citet{stansberry07}, however, these corrections are small in the MIPS
24~\micron\ band:  they are no more than 3\% over a range of power-law indices
($-$3 to 3) and do not exceed 5\% for blackbody sources above a temperature of
57~K.

\section{Sample and Predicted Fluxes}
\label{sec:sample}

The sample of stars used in \S~\ref{sec:calfactor} to derive the 24~\micron\
calibration factor covers a limited dynamic range and is too small to
probe statistically characteristics of the instrument behavior such as
linearity and the effects of sky brightness.  Furthermore, these stars are too
faint to be useful as calibrators for the MIPS 70~\micron\ band
\citep[discussed by][]{gordon07}.  In this section, we develop an expanded
sample of 24~\micron\ calibrators to support our characterization of the
performance of the instrument, which we discuss in \S~\ref{sec:checks}.  These
stars were selected to explore the dynamic range of the instrument within
reasonable integration times and to probe the effects of spectral type and
environment (primarily foreground/background level) on the calibration.

\subsection{Zero-point Conversions for Additional Data Sets}
\label{sec:zpconv}

We predict flux densities at 24~\micron\ for the expanded MIPS
calibration target list by extrapolating measurements made at near- and
mid-infrared wavelengths.  The scale factors for the 2MASS \citep{skrutskie06}
measurements of A and G stars made in the read-1 (the 51~ms exposure used
to measure sources that saturate the full 1.3~s exposure) data are given by
\citet{rieke07}, who find $K_s-[24]$ is 0 (by definition) and $0.045\pm0.011$
magnitudes, respectively.  The 2MASS read-1 magnitude limit
($\sim3.5$~magnitudes at $K_s$) is not low enough to constrain measurements
throughout the MIPS dynamic range, so we must incorporate additional data sets
into our flux predictions for the bright 24~\micron\ calibrators.  We discuss
here the validation of additional data sets and derive the scale factors
required to put them on the same system as the 2MASS read-1 measurements 
(and therefore the same system as the MIPS 24~\micron\ measurements),
allowing direct comparison to the results of \citet{rieke07}.

\subsubsection{2MASS Saturated-Source Magnitudes}

Below magnitudes of $\sim$~4.5, 4, and 3.5 at $J$, $H$, and $K_s$,
respectively, 2MASS data are saturated even in the read-1 measurements used to
constrain the calibration factor derived in \S~\ref{sec:calfactor}.
Measurements for saturated sources are reported by the 2MASS project and
are obtained by fitting to the radial profiles of the unsaturated portion of
the stellar image, but large uncertainties 20\%$-$30\% are associated
with these measurements.  To reduce the uncertainties associated with using
the saturated 2MASS magnitudes, we compute the offset between the saturated
and read-1 magnitudes.

\citet{kimeswenger05} presents photometry in the $J$ and $K_s$ bands for a
sample of $\sim600$ stars that overlaps the 2MASS read-1 and saturated-source
magnitude ranges.  This bright-star survey uses the same camera as the Deep
Near-Infrared Survey of the Southern Sky (DENIS) with the addition of
neutral-density filters, so we apply the DENIS transformations computed by
\citet{carpenter01} to put the magnitudes on the 2MASS system.  (This
step isn't strictly necessary since, as we discuss below, we are using these
measurements differentially.)  We compare the 2MASS measurements to those by
\citet{kimeswenger05} over different magnitude ranges to determine the offset
between the 2MASS read-1 and saturated-source magnitudes.  At faint levels
(well above the limits quoted above, to ensure that read-1 measurements were
not affected by any saturated sources), the 2MASS and the transformed
\citet{kimeswenger05} magnitudes are in excellent agreement: the
weighted average (rejecting points more than $3\sigma$ from the median) of
$J(2MASS)-J(Kimeswenger)$ between magnitudes 8.5 and 5.6 is $0.001\pm0.002$
magnitudes, and $K_s(2MASS)-K_s(Kimeswenger)$ between magnitudes 7.5 and 4.2
is $0.023\pm0.003$ magnitudes.  Below those magnitude limits, the offsets are
$J(2MASS)-J(Kimeswenger)=0.053\pm0.005$ magnitudes and
$K_s(2MASS)-K_s(Kimeswenger)=0.058\pm0.008$ magnitudes.  The difference
between the faint $J$ magnitudes is not significant, but we do include the
difference between the faint $K_s$ magnitudes in computing offsets
$(m_{read-1}-m_{saturated})$ of $-0.053$ magnitudes at $J$ and $-0.035$
magnitudes at $K_s$.  These offsets are added to saturated 2MASS magnitudes to
put them on the read-1 scale.  To further reduce the uncertainty, we average
the $J$ and $K_s$ magnitudes to compute a ``super''-$K_s$ magnitude, after
correcting $J$ to $K_s$ using the $J-K$ colors \citep[after correcting the
colors to the 2MASS system using the transformations computed
by][]{carpenter01} of stars tabulated by \citet{tokunaga00}.

\subsubsection{Johnson Photometry}

\citet{johnson66} measured $\sim650$ bright stars in the $J$ and $K$ bands.
This sample is large, homogeneously observed, and the measurements have
significantly smaller uncertainties than the saturated sources observed by
2MASS.  For example, the star HD~001013 was observed 91 times by
\citet{johnson66}.  After applying an airmass correction derived from the
data, we find a root-mean-square (RMS) deviation of 0.037 magnitudes in these
measurements.  We conservatively assign a $1\sigma$ uncertainty of 0.04
magnitudes to the photometry by \citet{johnson66}.

To compute the offset between the \citet{johnson66} magnitudes and 2MASS, we
transform the measurements to the 2MASS system using the equations derived by
\citet{carpenter01} for the \citet{koornneef83} system, which is most similar
to the system used by \citet{johnson66}.  The weighted average (rejecting
points more than $3\sigma$ from the median) of $J(2MASS)-J(transformed\
Johnson)$ is $0.028\pm0.007$ magnitudes and $K_s(2MASS)-K(transformed\
Johnson)$ is $0.004\pm0.005$ magnitudes.  The offset in the $K_s$ band is not
significant, but we apply a correction of 0.028 magnitudes to the measurements
at $J$.  In addition, since all the MIPS calibrators that were also measured
by \citet{johnson66} are well into the saturated 2MASS range, we apply the
same offsets applied to the saturated 2MASS measurements, for net corrections
of $-0.025$ magnitudes at $J$ and $-0.035$ magnitudes at $K_s$ (plus the color
transformation from Johnson to 2MASS).  Finally, we combine the $J$ and $K_s$
data to form super-$K_s$ as for the 2MASS data.

\subsubsection{IRAS Measurements}
\label{sec:iras_data}

Many of the bright MIPS calibrators were detected by IRAS at 12 and
25~\micron.  The MIPS 24~\micron\ band was demonstrated by \citet{rieke07} to
be linear within $\sim1$\% below 1~Jy, so we can use the overlap between MIPS
and IRAS measurements in this flux range to constrain the linearity of IRAS
and probe for effects of molecular absorptions in the 12~\micron\ band.  To
ensure that the IRAS measurements are on the same scale as the other
photometry we use to predict flux densities for our calibrators, we
empirically derive the ratio between the IRAS bands and the MIPS 24~\micron\
band below 1~Jy for the calibrator stars, then apply this result over the full
range of calibrator fluxes.

\begin{figure}
\plotone{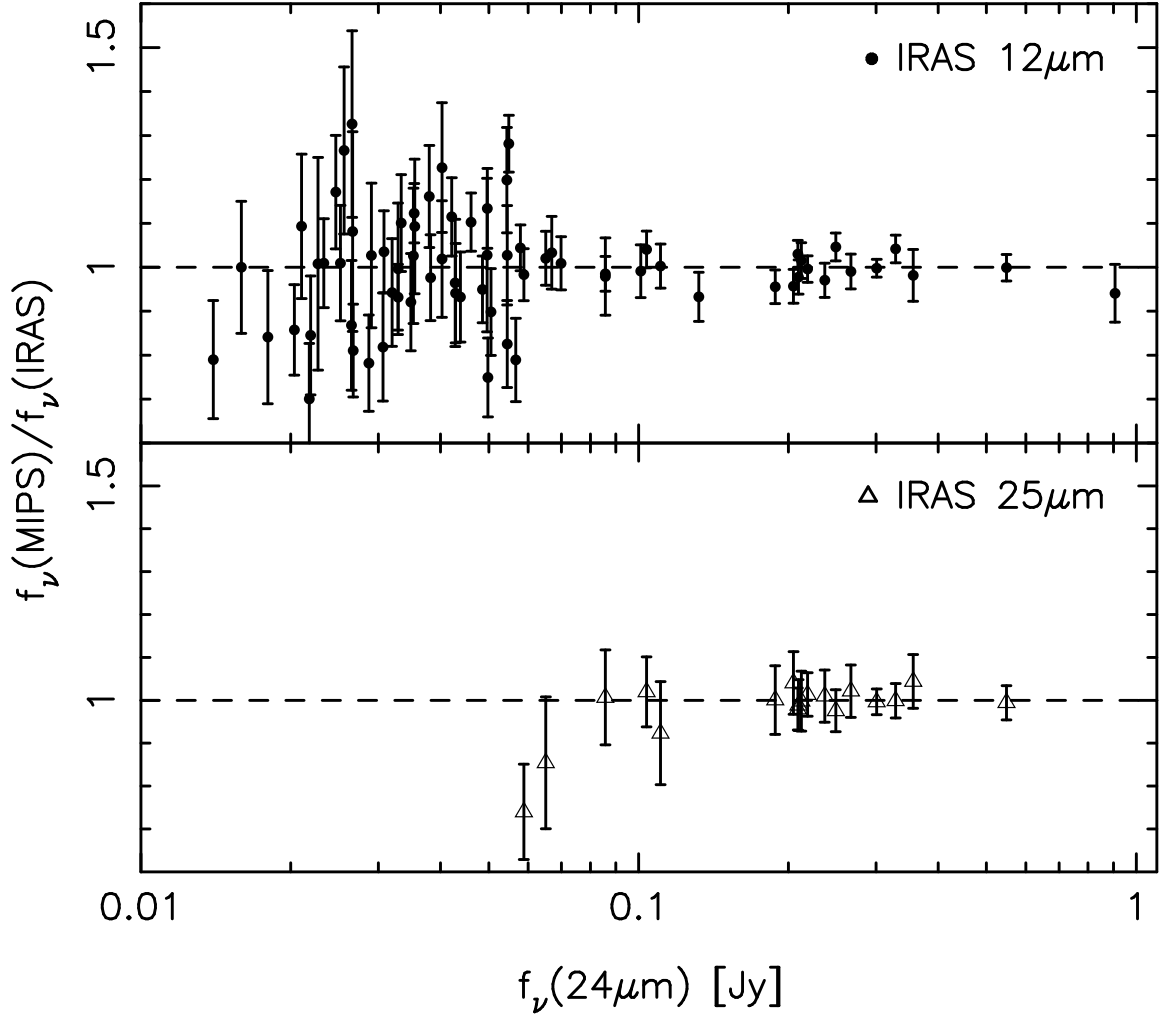}
\caption{The ratio of MIPS 24~\micron\ flux densities to IRAS 12 and
25~\micron\ flux densities (filled circles and open triangles, respectively)
as a function of 24~\micron\ flux density, normalized to the average ratios of
0.265 and 1.11, respectively.  The error bars represent the combined IRAS and
MIPS uncertainties.  A dashed line is drawn at a ratio of 1 as a
guide.\label{fig:iras-brightness}}
\end{figure}

\begin{figure}
\plotone{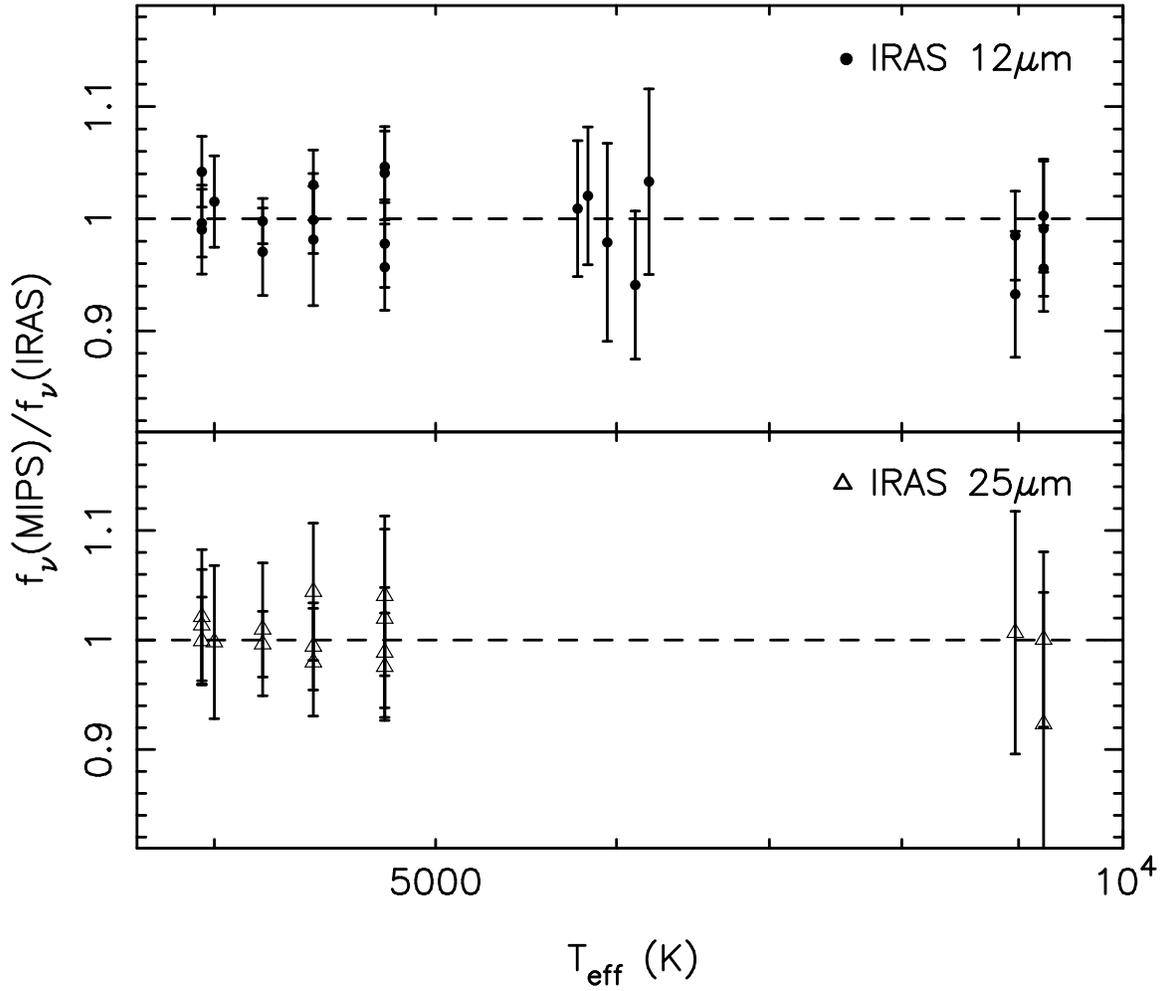}
\caption{The ratio of MIPS 24~\micron\ flux densities to IRAS 12 and
25~\micron\ flux densities (filled circles and open triangles, respectively)
as a function of spectral type, here quantified as the effective temperature
of the star.  The ratios have been normalized to the average ratios at 0.265
and 1.11 at 12 and 25~\micron.  The error bars represent the combined IRAS and
MIPS uncertainties.  A dashed line is drawn at a ratio of 1 as a
guide.\label{fig:iras-temperature}}
\end{figure}

We obtain IRAS measurements of the MIPS calibrators from the faint source
catalog (FSC; used because the improved sensitivity relative to the point
source catalog provides more overlap with the MIPS sample), using only
high-quality (quality flag ``3'') measurements.  We applied color corrections
appropriate to the temperature of each star to these observations and also
applied the corrections measured by \citet{rieke07}: 0.992 at 12~\micron\ and
0.980 at 25~\micron.  The ratio of the MIPS 24~\micron\ measurement to the
IRAS measurement, normalized to the average ratio, is plotted in
Figure~\ref{fig:iras-brightness}.  The scatter in the measurements becomes
obviously larger near the IRAS detection limit, around the equivalent
24~\micron\ flux density of 60 and 80~mJy at 12 and 25~\micron, respectively.
The slope of the ratio as a function of brightness above these limits is
consistent with 0 for both bands, indicating they are linear in this
brightness range.  The average value of
$f_{\nu}(24~\micron)/f_{\nu}(12~\micron)$ is 0.265 while
$f_{\nu}(24~\micron)/f_{\nu}(25~\micron)$ is 1.11.  Both values are very
similar to the values derived from Kurucz (1993) models of these stars, 0.266
and 1.11, respectively.  As shown in Figure~\ref{fig:iras-temperature},
$f_{\nu}(24~\micron)/f_{\nu}(12~\micron)$ has a mild dependence on temperature
(possibly the result of molecular absorptions in the 12~\micron\ band becoming
important in cool stars), so the ratio is better described as $0.276 -
1.94\times10^{-6}*T_{eff}[K]$.  No significant trend with temperature is
detected at 25~\micron.

The factors we derive in this section to convert the zero points of
saturated 2MASS magnitudes, \citet{johnson66} magnitudes, and IRAS 12~\micron\
and 25~\micron\ measurements to the system used by MIPS are summarized in
Table~\ref{tab:zpconv}.

\subsection{The Calibration of Cool Stars}

We compute an average super-$K_s-[24]$ (by applying the
aperture correction, 1.08, derived in \S~\ref{sec:photometry} and the
calibration factor derived in \S~\ref{sec:calfactor} to the measurements in
Table~\ref{tab:measurements}) color for the cool (K and M giants) stars in
our sample, which can be compared directly to the A and G star color computed
by \citet{rieke07}, modulo any offset (typically a fraction of a percent)
between super-$K_s$ and $K_s$.  The weighted average color of the K and M
stars in our sample (after rejecting points greater than $3\sigma$ from the
mean) is $0.104\pm0.006$ magnitudes.

\subsection{Predictions}
\label{sec:predictions}

We list the photometry used to constrain the flux density predictions, after
applying the corrections detailed above, in Table~\ref{tab:samp-pred}.  The
24~\micron\ flux densities for each source are the weighted average of the
predictions derived from the entries in that table.  We also list the
predicted flux densities and their uncertainties in Table~\ref{tab:samp-pred}.
We note that the predictions can be slightly different than those implicit in
Table~\ref{tab:cal-factor}, since we used super-K$_s$ in
Table~\ref{tab:samp-pred}  and K$_s$ was used
in Table~\ref{tab:cal-factor}, although the average difference is
insignificant:  0.002 $\pm$ 0.005 magnitudes.  We also include predicted
background levels, computed using the {\it Spitzer} Planning Observations
Tool\footnote{http://ssc.spitzer.caltech.edu/propkit/spot/} (SPOT); the
background listed is the average of the range when the target is visible, and
the uncertainty is half the difference between the extreme values.

\section{Checks On the 24~\micron\ Calibration}
\label{sec:checks}

In this section, we perform various checks on the 24~\micron\ calibration,
such as repeatability, linearity, and the effects of spectral type,
exposure time, and background.  Except for the repeatability check, we
compute a single calibration factor for each star by dividing the prediction
(Table~\ref{tab:samp-pred}) by the pixel area and the weighted average of all
measurements of the count rate for each star
(Table~\ref{tab:combined-measurements}).  The calibration factors derived from
some stars differ from the adopted calibration factor by more than
$5\sigma$---these stars were not used in the checks below.  As shown in
Table~\ref{tab:rejected}, most of the rejected stars show infrared fluxes
above the predictions (i.e., the calibration factor is low).

\subsection{Repeatability}

\begin{figure}
\plotone{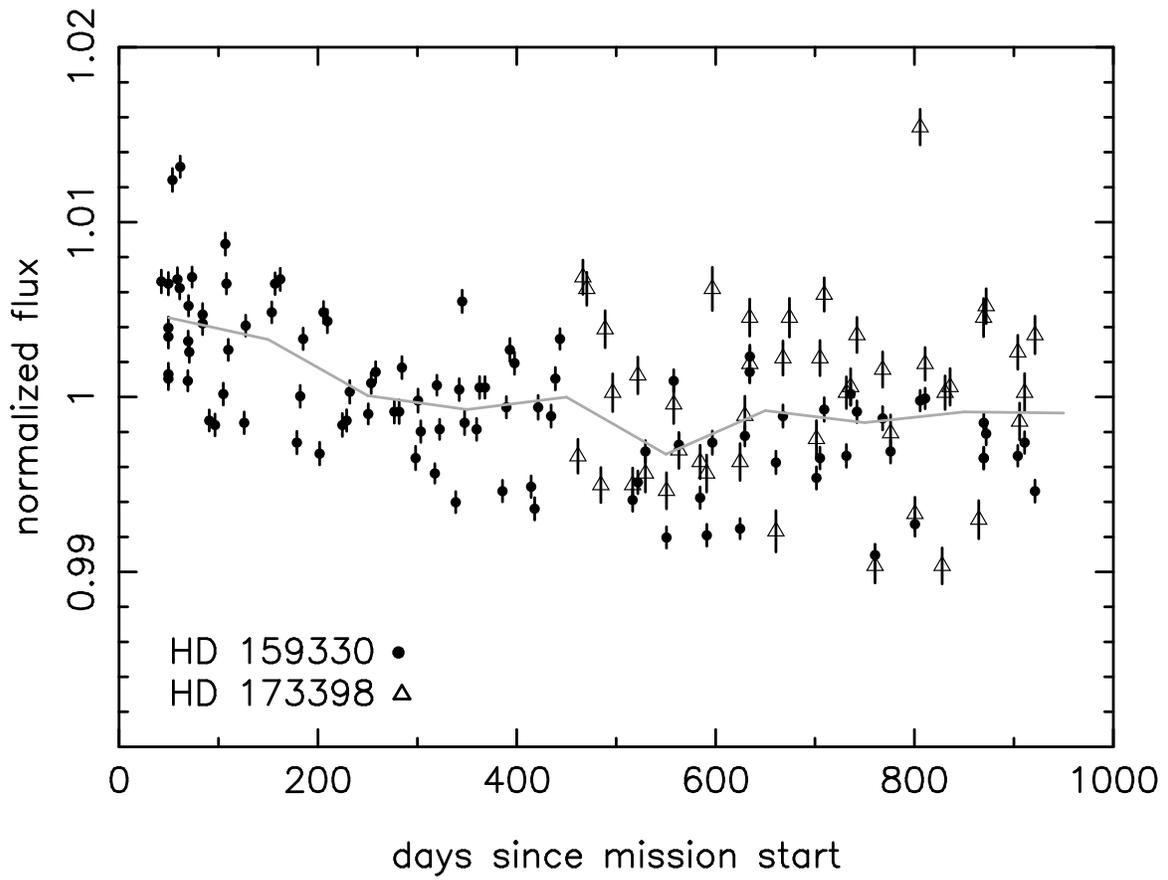}
\caption{Repeatability of 24~\micron\ photometry on two routinely-monitored
sources:  HD~159330 (filled circles) and HD~173398 (open triangles).  The gray
curve is a sigma-clipped average in 10 equally-spaced
bins.\label{fig:repeatability}}
\end{figure}

The primary routine calibrator for the MIPS 24~\micron\ channel is HD~159330,
a K2III star near the {\it Spitzer} continuous viewing zone (CVZ).  When
visible, this star is observed each time the instrument is turned on, to
monitor photometric stability and check for changes in the calibration.  A
second routine calibrator, the K0III star HD~173398, is also monitored to fill
in the gaps when HD~159330 is not visible.  We plot 100 measurements of
HD~159330 and 46 measurements of HD~173398 in Figure~\ref{fig:repeatability}.
The RMS scatter in the HD~159330 measurements is 0.4\% (compared to 0.7\%
in the SSC-pipeline-reduced data discussed in \S~\ref{sec:comparison}), while
the scatter in the HD~173398 measurements is 0.5\%.  A gradual reduction in
the instrument response of $\sim0.5$\% over the first 300 days is apparent in
the data and is the cause of some of the scatter computed for HD~159330.  As
this trend is insignificant compared to the uncertainty on the absolute
calibration (cf.  \S~\ref{sec:calfactor}), we have not attempted to correct
it.

\subsection{Linearity}

\begin{figure}
\plotone{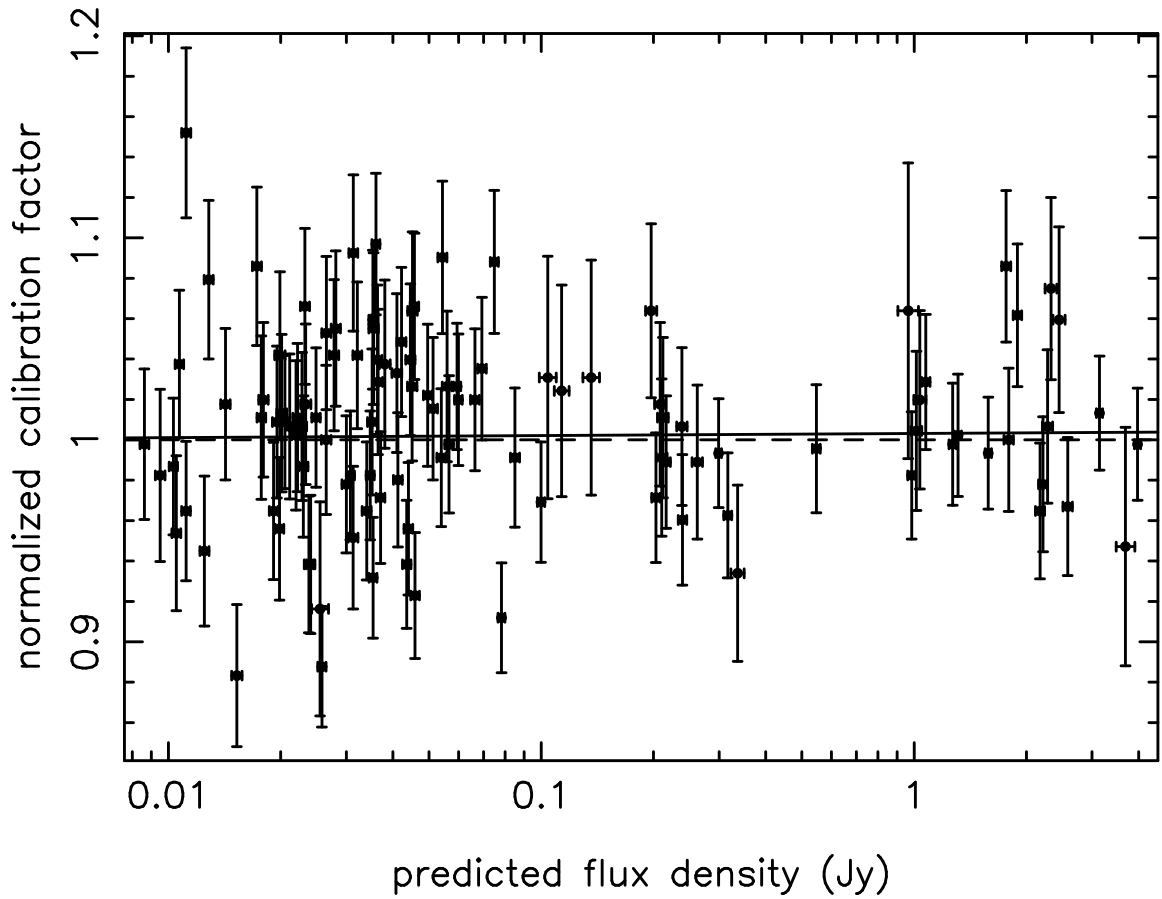}
\caption{24~\micron\ calibration factor normalized to the nominal calibration
factor (see \S~\ref{sec:calfactor}) as a function of predicted 24~\micron\
flux density.  The error bars are drawn from Tables~\ref{tab:samp-pred} and
\ref{tab:combined-measurements}.  The dashed line is drawn at 1 as a guide,
while the solid line is a linear least-squares fit to the data (see
\S~\ref{sec:checks}).  Points greater than 5$\sigma$ above or below the
nominal calibration factor have been not been included in the plot or the
fit.\label{fig:linearity}}
\end{figure}

We check for effects of flux nonlinearity in the calibration by comparing
calibration factors over a range of 460 in source brightness.  We plot the
calibration factors in Figure~\ref{fig:linearity}.  We find no significant
trend of calibration factor with source brightness---a least-squares fit to
the data shows a difference of only 0.3\% between 9~mJy and 4~Jy.  The
observed scatter of calibration factors is larger than can be explained by the
error bars.  The unaccounted-for scatter in the calibration factors is likely
due to systematic uncertainties on the flux predictions for the individual
stars, possibly from variability or small infrared excesses.

\subsection{Spectral Type}

As discussed in \S\ref{sec:sample}, we have derived calibration factors using
3 broad types of stars (hot dwarfs, solar analogs, and cool giants) to look
for systematic effects with stellar temperature.  The weighted average
calibration factor of each broad spectral type is $4.49\times10^{-2}$,
$4.62\times10^{-2}$, and $4.49\times10^{-2}$~MJy/sr/(DN/s) for 32 A, 37 G, and
25 K/M stars, respectively.  These values are all consistent with the adopted
calibration factor within the uncertainty (deviating by -1.1\%, 1.8\%, and
-1.1\%, respectively), but the differences may reflect real uncertainties in
the colors of different types of stars and in our treatment of saturated 2MASS
magnitudes.

\subsection{Exposure Time}
\label{sec:exptime}

A small subset of the calibrators (11 stars) was measured using 10- and
30-second exposures in addition to the 3-second exposures---the weighted
average counts per second from these stars is 1\% and 2\% higher using 10- and
30-second exposures, respectively.  We find that the residual images
in the 10- and 30-second exposures are roughly twice as bright as in the
3-second exposures, so it is likely that the excess is due to buildup of
residual charge during the longer exposures.

\subsection{Background}

\begin{figure}
\plotone{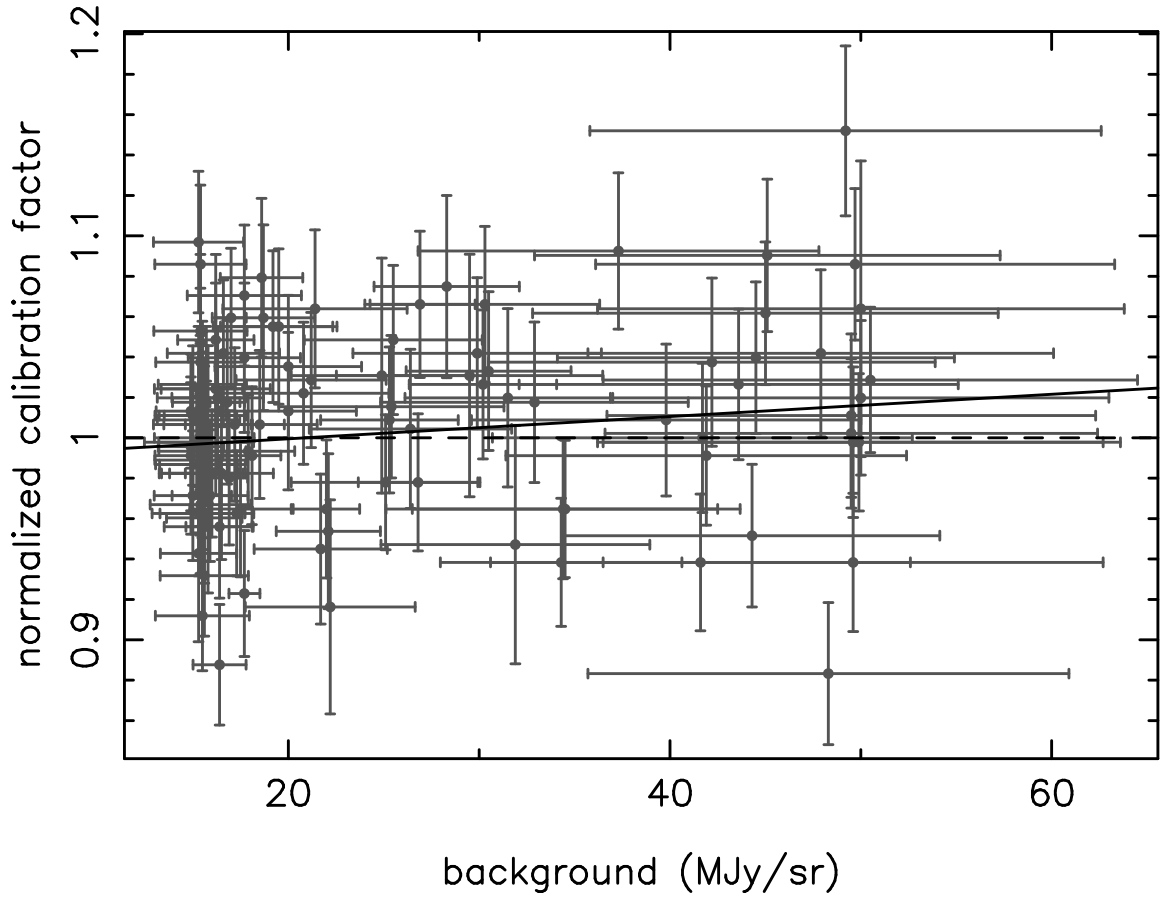}
\caption{24~\micron\ calibration factor normalized to the nominal calibration
factor (see \S~\ref{sec:calfactor}) as a function of predicted background.
The uncertainties represented by the error bars are drawn from
Tables~\ref{tab:samp-pred} and \ref{tab:combined-measurements}.  The dashed
line is drawn at 1 as a guide, while the solid line represents a linear
least-squares fit to the data (see \S~\ref{sec:checks}).  Points greater than
5$\sigma$ above or below the nominal calibration factor have been not been
included in the plot or the fit.\label{fig:cal_v_bkgd}}
\end{figure}

In general, astronomical sources of interest at 24~\micron\ will be observed
against a wide range of background levels, so we examine whether the
derived calibration factor depends on background.  Such an effect might be
expected due to background light scattering onto the detector or due to
systematic effects on the droop correction.  We find no significant
effect on the calibration over a factor of $\sim5$ in background, as shown in
Figure~\ref{fig:cal_v_bkgd}.  A least-squares fit to the data indicates a
slope of only marginal significance:
$5.4\times10^{-4}\pm2.9\times10^{-4}$~MJy/sr/(DN/s)~/~MJy/sr.

\subsection{Comparison to Another Infrared Calibration}

As discussed in detail by \citet{rieke07} and cited in \S~\ref{sec:iras_data},
the calibration presented here is based on an updated calibration system which
is offset by a small ($\sim2-3$\%) amount from other infrared calibrations
commonly in use.  For the convenience of the reader, we present a direct
comparison of our measurements to predictions in one of those systems, that
prepared by Cohen and collaborators.  Specifically, we compute predicted flux
densities at 24~\micron\ for the 10 ``template'' stars \citep[][and references
therein]{cohen99} observed by us by interpolating the values given by the
templates.  We adopt the model uncertainties provided by the templates.  We
take the ``observed'' values and uncertainties from
Table~\ref{tab:combined-measurements} and convert them to flux densities using
the calibration factor computed in \S~\ref{sec:calfactor}.  We present the
data used for and the results of this comparison in
Table~\ref{tab:cohen-comp}, where we reject the star HD020722 due to a
contaminating background source  at 24~\micron\ (cf.
Table~\ref{tab:rejected}).  The weighted average ratio of our measurements to
the predictions is $1.026\pm0.013$, i.e., measurements on the \citet{cohen99}
system can be converted to the system used by MIPS by multiplying by this
factor.  Much of this difference is due to the different fluxes adopted
for Vega at 10.6~\micron\ (35.03~Jy by \citet{rieke07} and 34.38~Jy by
\citet{cohen99}, which differ by 2\%).

\subsection{Extended-Source Calibration}

\begin{figure}
\plotone{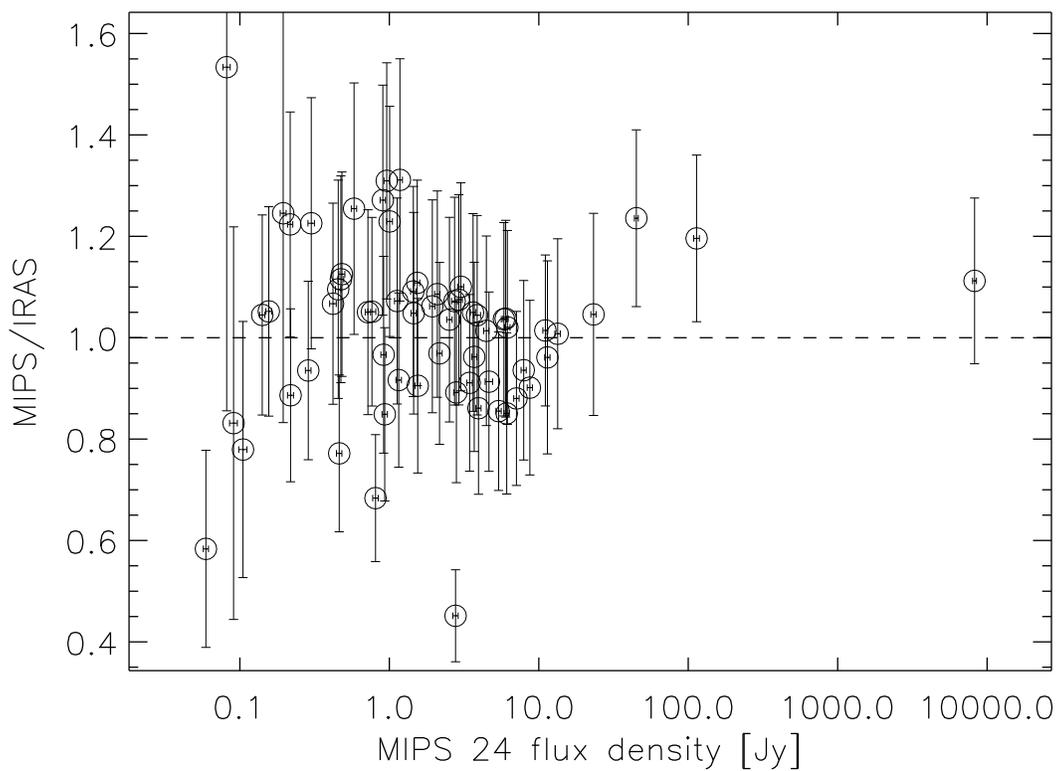}
\caption{Ratio of MIPS to IRAS measurements of extended sources as a function
of flux density measured at 24~\micron.  The error bars represent the combined
uncertainty on both measurements.  The dashed line is drawn at 1 as a guide.
\label{fig:extcheck}}
\end{figure}

As demonstrated in \S~\ref{sec:data}, the model PSF is a good match to the
data out to large radii, so we expect the extrapolation of the calibration to
infinite radii to be well understood.  As a check, we compare measurements of
extended sources (nearby galaxies) observed by the {\it Spitzer} legacy teams
SINGS \citep[{\it Spitzer} Infrared Nearby Galaxies Survey;][]{kennicutt03}
and SAGE \citep[Surveying the Agents of a Galaxy's Evolution;][]{meixner06}
and by guaranteed time observers \citep{hinz04,gordon06a,gordon06b} to IRAS
measurements.  We apply color corrections from \citet{beichman88} (IRAS) and
\citet{stansberry07} (MIPS) assuming a power-law spectrum fit to the
12/25~\micron\ or 24/70~\micron\ data to the IRAS and MIPS \citep[SINGS
measurements taken from][]{dale07} measurements, and also correct the MIPS
measurements to the calibration factor derived in \S~\ref{sec:calfactor}.  We
interpolate the IRAS 12~\micron\ and 25~\micron\ measurements to the effective
wavelength of the 24~\micron\ band, 23.675~\micron.  We compare the results
graphically in Figure~\ref{fig:extcheck}.  The weighted average ratio of MIPS
to IRAS measurements is 0.96, well within the 8\% combined uncertainty of both
instruments.

\section{Summary}
\label{sec:summary}

We discuss the flux calibration of the MIPS 24~\micron\ band, which is based
on stars.  We describe the data reduction and photometric procedures we use
for the calibration sources, which produce fluxes that are $1.6\pm0.6$\% lower
than those achieved by the automated pipeline at the {\it Spitzer} Science
Center.  We show that the calibration of the two imaging modes, photometry and
scan map, is consistent within 1\%.

We compute the calibration factor (the conversion from count rate to physical
units) for the MIPS 24~\micron\ band, using a sample of 22 A stars that has
been well measured and has been carefully vetted to exclude debris-disk
systems.  We find a value of $4.54\times10^{-2}$~MJy~sr$^{-1}$~(DN/s)$^{-1}$,
with an uncertainty of 2\%.  Based on this uncertainty and the difference
between the SSC pipeline and the DAT discussed in \S~\ref{sec:comparison} 
and summarized above, we recommend that a net uncertainty on the calibration
of 4\% is appropriate for general use.

We present a sample of 238 stars appropriate for use as MIPS flux
calibrators.  We have computed flux densities of these stars at the effective
wavelength of the 24~\micron\ band, 23.675~\micron.  We present 348
measurements of 141 of these stars, and combine those measurements with
the 24~\micron\ predictions to test various aspects of the calibration.  We
find that routine monitoring of a star near the {\it Spitzer} constant
viewing zone demonstrates that 24~\micron\ photometry with MIPS is repeatable
to 0.4\%.  The calibration is linear to 0.3\% over a range of $\sim460$ in
flux density, and there are no significant systematic effects on the
calibration due to spectral type, background, or angular extent of the source.

\acknowledgements

We thank John Carpenter for helpful discussions, especially regarding the
effect of exposure time on the measured count rate.  We would also
like to thank the anonymous referee, whose comments improved this paper.
This work is based on
observations made with the {\em Spitzer Space Telescope}, which is operated by
the Jet Propulsion Laboratory, California Institute of Technology under NASA
contract 1407.  This research has made use of the SIMBAD database, operated at
CDS, Strasbourg, France.  This publication makes use of data products from the
Two Micron All Sky Survey, which is a joint project of the University of
Massachusetts and the Infrared Processing and Analysis Center/California
Institute of Technology, funded by the National Aeronautics and Space
Administration and the National Science Foundation.  Support for this work was
provided by NASA through Contract Number \#1255094 issued by JPL/Caltech.

\clearpage

\begin{deluxetable}{lrlllll}
\tablewidth{0pt}
\tablecaption{24~\micron\ Aperture Correction Factors for a 10,000~K Blackbody.\tablenotemark{a}\label{tab:aperture-corrections}}
\tablehead {& & \multicolumn{5}{c}{background annulus}\\\cline{3-7}\colhead{description} & \colhead{radius} & \colhead{6-8} & \colhead{7-13} & \colhead{20-32} & \colhead{40-50} & \colhead{none}}
\startdata
half first dark ring & 3.5 & 2.78 & 2.80 & 2.57 & 2.56 & 2.56 \\
center of first dark ring & 7.0 & \nodata & 2.05 & 1.61 & 1.61 & 1.61 \\
outside first bright ring & 13.0 & \nodata & \nodata & 1.17 & 1.16 & 1.16 \\
center of second dark ring & 20.0 & \nodata & \nodata & 1.15 & 1.13 & 1.12 \\
outside second bright ring & 35.0 & \nodata & \nodata & \nodata & 1.08\tablenotemark{b} & 1.06 \\
\enddata
\tablenotetext{a}{All radii are measured in arcseconds.}
\tablenotetext{b}{Aperture correction used for calibration.}
\end{deluxetable}

\clearpage

\begin{longtable}{rrrrrr}
\tablewidth{0pt}
\tablecaption{24~\micron\ Measurements of MIPS flux calibrators.\label{tab:measurements}}
\tablehead {\colhead{}& \colhead{days since}& \colhead{}&
\colhead{DCE\tablenotemark{b} time}& \colhead{count rate}& \colhead{uncertainty}\\
\colhead{Name}& \colhead{mission start}& \colhead{AORKEY\tablenotemark{a}}& \colhead{(s)}& \colhead{(DN/s)}& \colhead{(DN/s)}}
\startdata
BD+621644 & 496.4 & 13071616 & 3 & 1.033e+04 & 5.19e+01 \\
HD000319 & 300.4 & 3972864 & 3 & 6.708e+03 & 5.00e+01 \\
HD001160 & 322.3 & 10090496 & 3 & 1.395e+03 & 1.56e+01 \\
HD001644 & 322.3 & 10090752 & 3 & 8.191e+03 & 5.69e+01 \\
HD002151 & 806.5 & 16276992 & 3 & 3.271e+05 & 1.23e+03 \\
HD002811 & 298.4 & 9940224 & 3 & 1.591e+03 & 3.35e+01 \\
HD008941 & 359.7 & 5414400 & 3 & 7.202e+03 & 3.46e+01 \\
HD009927 & 871.5 & 16619776 & 3 & 6.045e+05 & 2.45e+03 \\
\enddata
\tablenotetext{a}{The ``AORKEY,'' or Astronomical Observation Request Key,
is used by the {\it Spitzer} Science Center to uniquely identify the observation.}
\tablenotetext{b}{``DCE'' refers to ``Data Collection Event,'' in this
case an individual image.}
\end{longtable}

\clearpage

\begin{deluxetable}{rrrrrr}
\tablewidth{0pt}
\tablecaption{Data Used to Compute the 24~\micron\ Calibration Factor\label{tab:cal-factor}}
\tablehead {\colhead{}& \colhead{$K_s$\tablenotemark{a}}& \colhead{24\micron\ count rate}& \colhead{uncertainty}& \colhead{calibration factor}& \colhead{uncertainty}\\
\colhead{Name}& \colhead{(mag.)}& \colhead{(DN/s)}& \colhead{(DN/s)}& \colhead{(MJy/sr/[DN/s])}& \colhead{(MJy/sr/[DN/s])}}
\startdata
HD000319 & 5.479 & 6.708e+03 & 5.00e+01 & 4.513e$-$02 & 1.39e$-$03 \\
HD002811 & 7.057 & 1.591e+03 & 3.35e+01 & 4.448e$-$02 & 1.63e$-$03 \\
HD011413 & 5.422 & 7.866e+03 & 3.87e+01 & 4.056e$-$02 & 1.23e$-$03 \\
HD014943 & 5.439 & 7.180e+03 & 2.84e+01 & 4.374e$-$02 & 1.32e$-$03 \\
HD015646 & 6.411 & 2.997e+03 & 4.85e+01 & 4.281e$-$02 & 1.46e$-$03 \\
HD017254 & 5.877 & 4.529e+03 & 1.97e+01 & 4.632e$-$02 & 1.40e$-$03 \\
HD020888 & 5.691 & 5.519e+03 & 2.04e+01 & 4.512e$-$02 & 1.36e$-$03 \\
HD021981 & 5.526 & 6.089e+03 & 1.33e+01 & 4.761e$-$02 & 1.43e$-$03 \\
\enddata
\tablecomments{The average calibration factor is
$4.54\times10^{-2}$~MJy~sr$^{-1}$~(DN/s)$^{-1}$, to which we've assigned an
uncertainty of 2\% (see \S~\ref{sec:calfactor}).}
\tablenotetext{a}{$K_s=$[24] for the stars in this table \citep[see][]{rieke07}, all of which are between types A0 and A6.}
\end{deluxetable}

\clearpage

\begin{deluxetable}{ll}
\tablewidth{0pt}
\tablecaption{Zero-point Conversions\label{tab:zpconv}}
\tablehead{\colhead{Data Set} & \colhead{Conversion Factor}}
\startdata
2MASS saturated J & $-0.053$ mag. \\
2MASS saturated K$_s$ & $-0.035$ mag. \\
Johnson J & $-0.025$ mag. \\
Johnson K & $-0.035$ mag. \\
IRAS 12~\micron & $\times0.266$ \\
IRAS 25~\micron & $\times1.11$ \\
\enddata
\tablecomments{These factors convert measurements to 24~\micron\ magnitudes
(NIR) or fluxes (IRAS) (see \S~\ref{sec:zpconv}).}
\end{deluxetable}

\clearpage

\begin{longtable}{rrrrrrrrrrrr}
\tabletypesize{\scriptsize}
\tablecaption{MIPS Stellar Flux Calibrator Sample.\label{tab:samp-pred}}
\tablehead {& & & & & & & & \multicolumn{4}{c}{predictions}
\\ \cline{9-12} \colhead{} & \colhead{Spectral} &
\colhead{super-$K_s$\tablenotemark{a}}& \colhead{unc.}&
\colhead{$f_\nu(12\micron)$\tablenotemark{b}}& \colhead{unc.}&
\colhead{$f_\nu(25\micron)$\tablenotemark{b}}&
\colhead{unc.}& \colhead{$f_\nu(24\micron)$\tablenotemark{c}}& \colhead{unc.}&
\colhead{bkgd$_{24}$\tablenotemark{d}}& \colhead{unc.}\\
\colhead{Name}& \colhead{Type}& \colhead{(mag.)}& \colhead{(mag.)}& \colhead{(Jy)}& \colhead{(Jy)}& \colhead{(Jy)}& \colhead{(Jy)}& \colhead{(Jy)}& \colhead{(Jy)}& \colhead{(MJy/sr)}& \colhead{(MJy/sr)}}
\startdata
BD+621644 & K5 & 5.258 & 0.013 &  &  &  &  & 6.355e$-$02 & 1.50e$-$03 & 1.54e+01 & 2.38e+00 \\
HD000319 & A1V & 5.541 & 0.014 &  &  &  &  & 4.356e$-$02 & 1.14e$-$03 & 3.43e+01 & 6.33e+00 \\
HD001160 & A0 & 7.020 & 0.018 &  &  &  &  & 1.116e$-$02 & 3.17e$-$04 & 4.92e+01 & 1.34e+01 \\
HD001644 & K0III & 5.370 & 0.016 & 2.700e$-$01 & 3.24e$-$02 &  &  & 5.659e$-$02 & 1.52e$-$03 & 4.99e+01 & 1.37e+01 \\
HD001753 & F5V & 7.730 & 0.017 &  &  &  &  & 6.047e$-$03 & 1.83e$-$04 & 1.81e+01 & 2.37e+00 \\
HD002151 & G2IV & 1.307 & 0.193 & 8.471e+00 & 3.39e$-$01 & 2.099e+00 & 1.05e$-$01 & 2.279e+00 & 7.22e$-$02 & 1.72e+01 & 2.59e+00 \\
HD002261 & K0III & $-$0.143 & 0.223 & 3.866e+01 & 2.71e+00 & 9.149e+00 & 6.40e$-$01 & 1.010e+01 & 4.95e$-$01 & 2.27e+01 & 2.91e+00 \\
HD002811 & A3V & 7.086 & 0.017 &  &  &  &  & 1.050e$-$02 & 2.92e$-$04 & 2.21e+01 & 2.73e+00 \\
\enddata
\tablenotetext{a}{Super-$K_s$ is the weighted average of $J$ (transformed to
$K_s$) and $K_s$
(see \S~\ref{sec:zpconv}).}
\tablenotetext{b}{Taken from the IRAS Faint Source Catalog, and modified
as discussed in \S~\ref{sec:iras_data}.}
\tablenotetext{c}{These flux densities apply to the effective wavelength of
the 24~\micron\ band, 23.675~\micron.}
\tablenotetext{d}{See \S~\ref{sec:predictions}.}
\end{longtable}

\clearpage

\begin{longtable}{rrrrrr}
\tabletypesize{\small}
\tablewidth{0pt}
\tablecaption{Combined Measurements and Calibration Factors\label{tab:combined-measurements}}
\tablehead {\colhead{}& \colhead{number of}& \colhead{24\micron\ count
rate\tablenotemark{a}}& \colhead{uncertainty}& \colhead{calibration factor}& \colhead{uncertainty}\\
\colhead{Name}& \colhead{measurements}& \colhead{(DN/s)}& \colhead{(DN/s)}& \colhead{(MJy/sr/[DN/s])}& \colhead{(MJy/sr/[DN/s])}}
\startdata
BD+621644 & 1 & 1.033e+04 & 5.19e+01 & 4.037e$-$02 & 9.72e$-$04 \\
HD000319 & 1 & 6.708e+03 & 5.00e+01 & 4.262e$-$02 & 1.16e$-$03 \\
HD001160 & 1 & 1.395e+03 & 1.56e+01 & 5.250e$-$02 & 1.60e$-$03 \\
HD001644 & 1 & 8.191e+03 & 5.69e+01 & 4.534e$-$02 & 1.25e$-$03 \\
HD002151 & 1 & 3.271e+05 & 1.23e+03 & 4.573e$-$02 & 1.46e$-$03 \\
HD002811 & 1 & 1.591e+03 & 3.35e+01 & 4.331e$-$02 & 1.51e$-$03 \\
HD008941 & 1 & 7.202e+03 & 3.46e+01 & 4.948e$-$02 & 1.40e$-$03 \\
HD009927 & 1 & 6.045e+05 & 2.45e+03 & 4.539e$-$02 & 1.17e$-$03 \\
\enddata
\tablenotetext{a}{The count rates in this table can be converted to Jy by multiplying by the product of the aperture correction, calibration factor, and pixel area discussed in the text, or a factor of $6.92\times10^{-6}$.}
\end{longtable}

\clearpage

\begin{deluxetable}{ll}
\tablewidth{0pt}
\tablecaption{Stars Rejected as 24~\micron\ Calibrators.\label{tab:rejected}}
\tablehead {\colhead{Name} & \colhead{Reason for Rejection}}
\startdata
BD+621644 & calibration factor 12\% low \\
HD011413 & calibration factor 18\% low\tablenotemark{a} \\
HD020722 & double source\tablenotemark{b} \\
HD046190 & calibration factor 29\% low\tablenotemark{c} \\
HD047332 & calibration factor 18\% low\tablenotemark{a} \\
HD065517 & calibration factor 16\% low\tablenotemark{a} \\
HD092845 & double star \\
HD102647 & calibration factor 29\% low\tablenotemark{c} \\
HD109612 & bright, complex background \\
HD109866 & bright, complex background \\
HD115780 & bright, complex background \\
HD119545 & bright, complex background \\
HD129655 & calibration factor 19\% low\tablenotemark{a} \\
HD131986 & bright, complex background and nearby contaminating sources \\
HD132417 & bright, complex background and nearby contaminating sources \\
HD132439 & bright, complex background \\
HD139698 & bright, complex background \\
HD158460 & calibration factor 21\% low\tablenotemark{c} \\
HD165459 & calibration factor 61\% low\tablenotemark{c} \\
HD201941 & calibration factor 26\% low\tablenotemark{d} \\
SAO9310 & calibration factor 20\% high \\
\enddata
\tablenotetext{a}{NIR measurements indicate this star is reddened, so the
predicted flux is likely low.}
\tablenotetext{b}{There is a 24\micron\ source 14\arcsec\ to the north of this
star.  SIMBAD does not indicate the star is part of a multiple system and the
{\it Spitzer} Planning and Observation Tool (SPOT) does not indicate any
asteroids in the field, so the source is likely a background galaxy.}
\tablenotetext{c}{\citet{su06} find this source has a debris disk.}
\tablenotetext{d}{This star is not known to be part of a multiple system and no
asteroids were expected in the field, so this star likely has an infrared
excess.}
\end{deluxetable}

\clearpage

\begin{deluxetable}{rrrrrrr}
\tablewidth{0pt}
\tablecaption{24~\micron\ Observations of Cohen et al. Template Stars\label{tab:cohen-comp}}
\tablehead { & \multicolumn{2}{c}{observed} &
\multicolumn{2}{c}{template prediction} \\
\colhead{} & \colhead{$f_\nu(24\micron)$} & \colhead{unc.} & \colhead{$f_\nu(24\micron)$} & \colhead{unc.} & \colhead{observed /} & \colhead{}\\
\colhead{Name} & \colhead{(Jy)} & \colhead{(Jy)} & \colhead{(Jy)} & \colhead{(Jy)} & \colhead{template} & \colhead{unc.}}
\startdata
HD032831 & 1.592e+00 & 5.58e$-$03 & 1.560e+00 & 5.99e$-$02 & 1.020 & 0.039 \\
HD053501 & 1.271e+00 & 2.57e$-$03 & 1.238e+00 & 1.08e$-$01 & 1.027 & 0.089 \\
HD163588 & 3.098e+00 & 9.27e$-$03 & 3.018e+00 & 1.06e$-$01 & 1.027 & 0.036 \\
HD189276 & 2.668e+00 & 7.06e$-$03 & 2.559e+00 & 8.31e$-$02 & 1.043 & 0.034 \\
HD138265 & 1.000e+00 & 2.93e$-$03 & 9.507e$-$01 & 3.25e$-$02 & 1.052 & 0.036 \\
HD036167 & 3.893e+00 & 1.54e$-$02 & 3.800e+00 & 1.27e$-$01 & 1.024 & 0.035 \\
HD170693 & 1.283e+00 & 1.48e$-$04 & 1.291e+00 & 4.57e$-$02 & 0.994 & 0.035 \\
HD009927 & 4.182e+00 & 1.69e$-$02 & 3.908e+00 & 2.94e$-$01 & 1.070 & 0.081 \\
HD134493 & 2.045e$-$01 & 8.58e$-$04 & 2.017e$-$01 & 6.90e$-$03 & 1.014 & 0.035 \\
\enddata
\tablecomments{The star HD020722 has a template and was observed at
24~\micron, but was not included in this comparison due to a nearby source at
24~\micron\ (cf. Table~\ref{tab:rejected}).}
\end{deluxetable}

\clearpage

\clearpage

\clearpage

\clearpage

\clearpage

\clearpage

\clearpage

\clearpage

\clearpage

\clearpage

\end{document}